\documentclass[preprint,preprintnumbers,superscriptaddress,amsmath]{revtex4}
\usepackage{url}
\usepackage{epsfig}
\usepackage{psfrag}
\usepackage[psamsfonts]{amssymb}
\usepackage{amsmath}
\usepackage{indentfirst}
\usepackage{amssymb}
\usepackage{wrapfig}
\newcommand{\bs}{\boldsymbol}

\begin{document}

\title{Generation-by-Generation Dissection of the Response Function\\ in Long Memory Epidemic Processes}
\author{A. Saichev}
\affiliation{Department of Management, Technology and Economics,
ETH Zurich, Kreuzplatz 5, CH-8032 Zurich, Switzerland}
\affiliation{Mathematical Department,
Nizhny Novgorod State University, Gagarin prosp. 23,
Nizhny Novgorod, 603950, Russia}
\email{saichev@hotmail.com,dsornette@ethz.ch}

\author{D. Sornette}
\affiliation{Department of Management, Technology and Economics,
ETH Zurich, Kreuzplatz 5, CH-8032 Zurich, Switzerland}

\begin{abstract}
 In a number of natural and social systems, the response to an exogenous shock
 relaxes back to the average level according to a long-memory kernel $\sim 1/t^{1+\theta}$
 with $0 \leq \theta <1$. In the presence of an epidemic-like process of triggered
 shocks developing in a cascade of generations at or close to criticality,
 this ``bare'' kernel is renormalized into an even slower decaying response
 function  $\sim 1/t^{1-\theta}$. Surprisingly, this means that
 the shorter the memory of the bare kernel (the larger $1+\theta$),
 the longer the memory of the response function (the smaller $1-\theta$).
 Here, we present a detailed investigation of this paradoxical behavior
based on a generation-by-generation decomposition of the total response function, the use of Laplace
transforms and of ``anomalous'' scaling arguments. The paradox is explained by
the fact that the number of triggered
generations grows anomalously with time at $\sim t^\theta$ so that the contributions
of active generations up to time $t$ more than compensate the shorter memory
associated with a larger exponent $\theta$.  This anomalous scaling results fundamentally from
the property that the expected waiting time is infinite for $0 \leq \theta \leq 1$.
The techniques developed here are also applied to the case $\theta >1$ and we find 
in this case that
the total renormalized response is a {\bf constant} for $t < 1/(1-n)$ followed by a cross-over to
$\sim 1/t^{1+\theta}$ for $t \gg 1/(1-n)$.
\end{abstract}

\date{\today}

\maketitle

\section{Introduction}

Many systems in the natural and social worlds are characterized by activities whose level
$A(t)$ at some time $t$ is a function of its past levels $\{A(\tau), ~{\rm for}~0 \leq \tau <t\}$.
This can be described by the generic integral equation
\begin{equation}
\label{ainteq}
A(t) = f(t) + n \int_0^t A(\tau) \Phi(t-\tau) d\tau~ ,
\end{equation}
where $f(t)$ is some source or perturbation term whose impact is instantaneous.
The second integral term describes the propagation of past activity levels $A(\tau)$
to the present time $t$ mediated by the kernel $\Phi(t-\tau)$. The summation describes
that all past activities have an impact in the present activity level, but with a weaker and
weaker weight $\Phi(t-\tau)$ as they recede more in the past.
The ``bare'' kernel function must satisfy the normalization condition
\begin{equation}\label{normphi}
\int_0^\infty \Phi(t) dt=1~.
\end{equation}
Finally, the parameter $n$, which is in  $(0,1)$ to ensure the absence of explosive solutions,
describes the relative strength of triggering of future activity by past activity, as will become clear in the sequel.
As equation (\ref{ainteq}) can be obtained as the statistical average of a large class
of epidemic branching models \cite{DVJ2007,HSbasic02}, it is natural to refer to $n$ as the ``branching ratio''.

We are interested in the class of systems for which the kernel $\Phi(t)$ expresses
the existence of a long memory and, for the sake of concreteness, our calculations
will use the specific form
\begin{equation}
\label{powkern0}
\Phi(t) = {\theta \varrho^\theta \over (t+\varrho)^{\theta+1} }~ ,
\end{equation}
corresponding to $\Phi(t) \sim 1/t^{1+\theta}$ at long times.

Expression (\ref{ainteq}) with (\ref{powkern0}) corresponds to a mean-field or statistical averaged
description of the dynamics of many systems \cite{sornette2005origins}, such as the following examples.
Present seismicity is in large part triggered by past seismicity over long time scales described by the Omori law \cite{Felzer02,HSbasic02,Freed05,SW05}, $A(t)$ being the seismic rate in a given region above some
magnitude threshold. Commercial and social successes  have been shown to promote success over very long time scales \cite{sornette2004amazon,deschatres2005amazon,Crane_YouTube}. The activity $A(t)$ here corresponds
to the number of products sold or the number of downloads, views, attendence and so on, per unit time. Past financial volatility
has a very long influence on future volatilty, leading to bursty intermittent behaviors also characterized
by long-memory power law decaying kernels \cite{Cont01,CalvetFisher08,MRWvol}. Here the activity $A(t)$
is simply a measure of the financial volatility.

 The solution of equation \eqref{ainteq} has the form
\begin{equation}
\label{generalsolAthrures}
A(t) = f(t) +  n\int_0^t R(t-\tau) f(\tau) d\tau~ ,
\end{equation}
where $R(t)$ is the \emph{resolvent}, also known at the renormalized kernel
or response function, satisfying the equation
\begin{equation}\label{resolventeq}
R(t) = \Phi(t) + n \int_0^t \Phi(t-\tau) R(\tau) d\tau~ .
\end{equation}
The name ``response function'' refers to the fact that, if
$f(t)=M \delta(t-t_0)$ is a delta function describing a sudden impulse of amplitude $M$ applied to the system
at some time $t_0$ then,
\begin{equation}\label{delares}
A(t) = M ~n ~R(t-t_0)
\end{equation}
for any $t>t_0$. The resolvent $R(t)$ thus describes the response of the system to an impulsive perturbation.

The focus of the present paper is to analyze the properties of the response function $R(t)$
by addressing two specific questions.
\begin{itemize}
\item PARADOX 1: {\it at criticality, the shorter the memory of the bare kernel, the longer the memory of the response function!}

For the class of systems with long memory with $0 \leq \theta <1$, starting with Montroll and Scher \cite{Montroll12}, a number of authors
have shown that
\begin{equation}
R(t) \sim 1/t^{1-\theta}~~~{\rm for}~n=1~~~ {\rm or ~if} ~n<1, ~{\rm for} ~t<t^*~,
\label{resostgws}
\end{equation}
where $t^* = \varrho ~\left({n~\Gamma(1-\theta) \over |1-n|}\right)^{1/\theta}$. For
$t>t^*$, $R(t) \sim \Phi(t) \sim 1/t^{1+\theta}$ \cite{Scher,SS99,HSbasic02} (see
in particular \cite{HSPRE02} for a synthesis).
The value $n=1$ corresponds to the critical regime.
While the derivation of this result is rather straightforward when using the Laplace
transform operator as we recall below, this result is paradoxical. Indeed, at face value,
it means that the larger $\theta$ is, the shorter is the memory encoded by the bare kernel
$\Phi(t)$, and the longer is the memory described by the response function $R(t)$.
A first goal of this paper is to solve this paradox by a detailed analysis
of the role played by the cascade of triggering events intrinsically embodied in equation (\ref{ainteq})
for $n=1$.

\item PROBLEM 2: {\it power law exponent of the resolvent for $1< \theta <2$ at criticality ($n=1$)}.

Recently, an empirical example of the regime where $1< \theta <2$ has been discovered
in the humanitarian response to the destruction brought by the tsunami generated by the Sumatra earthquake of December 26, 2004, as measured by donations \cite{Tsunami_donation}. The data suggests that $n<1$ so that
the observed response is $R(t) \sim \Phi(t) \sim 1/t^{1+\theta}$, i.e., the effect of multiple triggering
occurring for $n \simeq 1$ did not occurred in this episode. Here, we ask what would be the
response function of a such a system in which the bare kernel is of the form (\ref{powkern0})
with $1< \theta <2$ if an epidemic cascade characterized by $n \simeq 1$ occurred. Clearly, the solution (\ref{resostgws}) cannot apply for $1< \theta <2$, since it would lead to a growth of the response function with time! The solution for this regime that we provide below  further illuminates the solution of Paradox 1.
\end{itemize}

In a nutshell, the solution of the paradox developed in the sequel of this
paper is based on the decomposition of the total activity
as the sum over a time-varying number $K(t) \sim t^\theta$ of generations which have been activited
until time $t$. Since each generation $k$, with $1 \leq k \leq K(t)$, contributes
to the total activity with an amplitude which is proportional to
$\sim k n^k /t^{1+\theta}$, the total activity is therefore $\sim {1 \over t^{1+\theta}} \times \left( \sum_{k=1}^{K(t)} k n^k \right) \sim [K(t)]^2 / t^{1+\theta} \sim t^{2\theta} / t^{1+\theta} \sim 1 /t^{1-\theta}$. Therefore, the larger
$\theta$ is, the shorter the memory of $1 / t^{1+\theta}$, but the faster growing is the number
of generations that are triggered up to time $t$.
Thus, the resulting slower decaying renormalized
response function $\sim 1 /t^{1-\theta}$  results from the fact that the number of triggered
generations grows sufficiently fast so as to more than compensate the shorter memory
associated with a larger $\theta$.  This anomalous scaling results fundamentally from
the property that the expected waiting time is infinite for $0 \leq \theta \leq 1$.
Adaptation of this reasoning to the case $1 < \theta$ implies that the renormalized
activity is a {\bf constant} for $t< 1/(1-n)$ followed by a cross-over to
$\sim 1/t^{1+\theta}$ for $t \gg 1/(1-n)$.

The rest of the paper develops the derivations of these results and is organized in the
three following sections.
The next section 2 constructs the mathematical building blocks used in the subsequent
sections. In particular, subsection \ref{thoggqw} summarizes the main results obtained
using the Laplace transform applied to expression (\ref{resolventeq}). Section 3
presents a detailed derivation of the solution of Paradox 1 in terms of a
generation-by-generation decomposition. Specifically, subsection \ref{theyt4ns}
dissects the two key contributions in the global activity and subsection \ref{intuitionparadox}
presents the intuitive derivation of Paradox 1 based on scaling arguments for long waiting times.
Section 4 applies the same approach in terms of the generation-by-generation decomposition
for the case $1 < \theta <2$.

\section{Mathematical building blocks}

\subsection{General relations obtained by using the Laplace transform operator}

The standard way to solve the integral equation \eqref{resolventeq} is to apply the Laplace transform.
This transforms the integral equation into an algebraic one
\begin{equation}\label{algeqlaptr}
\tilde{R}(s) = \tilde{\Phi}(s) + n \tilde{\Phi}(s) \tilde{R}(s)~ ,
\end{equation}
where the tilde denotes that the corresponding function is the Laplace image of the original function.
For instance, the Laplace image of the \emph{kernel} $\Phi(t)$ is
\begin{equation}
\tilde{\Phi}(s) = \int_0^\infty \Phi(t) e^{-s t} dt ~ .
\end{equation}
It follows from \eqref{algeqlaptr} that the Laplace image of the resolvent is given by
\begin{equation}\label{lapimresolv}
\tilde{R}(s) = { \tilde{\Phi}(s) \over 1- n \tilde{\Phi}(s)} = {1 \over n} \sum_{k=1}^\infty n^k \tilde{\Phi}^k(s) ~ .
\end{equation}
Accordingly, one can represent the solution $A(t)$ of \eqref{delares} in the form
\begin{equation}\label{ainformofser}
A(t) = \sum_{k=1}^\infty A_k(t)~ ,
\end{equation}
where
\begin{equation}
A_k(t) = M n^k \Phi_k(t)
\end{equation}
and $\Phi_k(t)$ is the inverse Laplace image of $\tilde{\Phi}^k(s)$.

The series \eqref{ainformofser} has a transparent meaning, when interpreted in the context of
epidemic processes. Interpreting $f(t)$ defined in \eqref{generalsolAthrures} as some ancestor event of
amplitude $M$, then $A(t)$ is the mean birth rate of its offsprings. Correspondingly, the $k$-th term $A_k(t)$ in the series \eqref{ainformofser} is the mean birth rate of those offsprings of the $k$-th generation. In this context, $n$ is the critical parameter of the corresponding branching process. In what follows, it is convenient to use the terminology of
branching processes in order to describe the characteristic properties of the resolvent $R(t)$ and the corresponding solution $A(t)$ of the integral equation \eqref{ainteq}. For simplicity, but without loss of generality, we put $M=1$.

\subsection{Scaled kernel and resolvent}

Specific calculations will be performed with the form (\ref{powkern0}) of the kernel. But,
whenever possible, we will keep the discussion as general as possible.  In particular, we
will consider the
general class of kernels $\Phi(t)$ of the integral equation \eqref{ainteq} which has the form
\begin{equation}\label{selfsimkernel}
\Phi(t) = {1 \over \varrho}\, \varphi\left({t \over \varrho} \right)~ ,
\end{equation}
where $\varrho>0$ is some unique characteristic time scale, while the kernel $\varphi(x)$ is some given function of its dimensionless argument, satisfying to normalization condition
\begin{equation}\label{normcondvarphi}
\int_0^\infty \varphi(x) dx =1 ~ .
\end{equation}
For the choice (\ref{powkern0}), we have
\begin{equation}
\varphi(x) = {\theta \over (x+1)^{\theta+1}}~.
\label{powkern}
\end{equation}
We will restrict our study to the case of kernels possessing the power asymptotics
$\Phi(t) \sim t^{-\theta-1}$ for $t\to\infty$, with $0 < \theta <2$. The case $\theta=1$ requires
a special treatment. We will not present it for the sake of conciseness, while it is clear that the method
presented below allows one to easily provide the needed detailed description in this case. We note that the main
scaling laws obtained below remain valid for $\theta=1$, while some details and corrections
to scaling differ.

For the family of kernels given by (\ref{selfsimkernel}) with a single characteristic scale
$\varrho$, the resolvent $R(t)$ and the mean activity $A(t)$ can be represented in forms analogous to \eqref{selfsimkernel}:
\begin{equation}\label{selfsimresolvent}
R(t) = {1 \over \varrho } \mathcal{R} \left( { t \over \varrho}, n \right) ~ , \quad A(t) = {1 \over \varrho } \mathcal{A} \left( { t \over \varrho}, n \right) ~ , \quad \mathcal{A}(x,n) = n\mathcal{R}(x,n)~ .
\end{equation}
The Laplace images of the resolvent $\mathcal{R}(x,n)$ and of the mean activity $\mathcal{A}(x,n)$ are given by
\begin{equation}\label{Lapimmothres}
\mathcal{\tilde{R}}(y,n) = {\tilde{\varphi}(y) \over 1 - n \tilde{\varphi}(y)} = {1 \over n} \sum_{k=1}^\infty n^k \tilde{\varphi}^k (y)~ , \quad \mathcal{\tilde{A}}(y,n) = \sum_{k=1}^\infty n^k \tilde{\varphi}^k (y)~ ,
\end{equation}
where
\begin{equation}
\tilde{\varphi}(y) = \int_0^\infty \varphi(x) e^{-xy} dx
\end{equation}
is the Laplace image of the bare kernel $\varphi(x)$.
In particular, the Laplace image of the kernel given by \eqref{powkern} is equal to
\begin{equation}\label{powermothkernlapim}
\tilde{\Phi}(s) = \theta\, (\varrho s)^\theta\, e^{\varrho s}\, \Gamma(-\theta,\varrho s)~ , \quad \Rightarrow \quad \tilde{\varphi}(y) = \theta y^\theta e^y \Gamma(-\theta, y)~.
\end{equation}

The asymptotic power law of the kernel $\varphi(x)\sim x^{-1-\theta}$ ($x\to \infty$) leads to
the following asymptotics for the Laplace image $\tilde{\varphi}(y)$ for small $y$ values:
\begin{equation}\label{genasymplapim}
\tilde{\varphi}(y) \simeq 1 +\alpha y - \beta y^\theta~ , \qquad |y|\ll 1~ .
\end{equation}
For the particular case \eqref{powkern}, we obtain
\begin{equation}\label{mothkernaprser}
\tilde{\varphi}(y) \simeq 1+ {y \over 1-\theta } - y^\theta\, \Gamma(1-\theta)~ , \qquad |y|\ll 1~ .
\end{equation}
In this case
\begin{equation}\label{albesigns}
\alpha \equiv \alpha(\theta) = {1 \over 1-\theta} ~ , \qquad \beta\equiv \beta(\theta) = \Gamma(1-\theta)~ .
\end{equation}
Notice that $\alpha(\theta)$ and $\beta(\theta)$ change sign as $\theta$ crosses the value $\theta=1$. Specifically, $\alpha(\theta)$ and $\beta(\theta)$ are negative for $\theta\in(1,2)$ and positive for $\theta\in(0,1)$. These signs are the consequence of the fact that $\Phi(t)$ given by \eqref{powkern} is one-sided, i.e., identically equal to zero for any $t<0$. In the following, we will thus
consider the general asymptotic expression \eqref{genasymplapim} with coefficients  $\alpha$ and $\beta$ of the same sign.

\subsection{Direct derivation of the resolvent $R(t)$ \label{thoggqw}}

Substituting (\ref{genasymplapim}) into (\ref{lapimresolv}) and using its first equality leads to
 \begin{equation}\label{resolvasymplaim}
\mathcal{\tilde{R}}(y,n) \simeq {1 \over q -\alpha y +\beta y^\theta} ~ , \qquad q= 1-n ~ .
\end{equation}

\subsubsection{Case $0<\theta<1$}

In this case, the term $\alpha y$ can be neglected in \eqref{resolvasymplaim}, which becomes
\begin{equation}\label{resthleson}
\mathcal{\tilde{R}}(y,n) \simeq {1 \over q +|\beta| y^\theta}~ .
\end{equation}
We have intentionally replaced $\beta$ by $|\beta|$ to stress the important fact that, for $\theta\in(0,1)$, the parameter $\beta$ is positive as can be checked from its explicit value given in (\ref{albesigns}).
For very small $y$, expression (\ref{resthleson}) can be further expanded into the following asymptotic relation
\begin{equation}\label{resthlesonwtg}
\mathcal{\tilde{R}}(y,n) \simeq {1 \over q}  -{|\beta| \over q^2}~ y^\theta~ , \qquad {|\beta| \over q} ~ |y|^\theta \ll 1~ .
\end{equation}
Using the standard correspondence between functions and their Laplace transforms
\begin{equation}\label{powerxyrel}
y^\theta \quad \mapsto \quad {x^{-\theta -1} \over \Gamma(-\theta)} ~ , \qquad 1 \quad \mapsto \quad \delta(x) ~ ,
\end{equation}
we obtain the asymptotic time dependence of the resolvent:
\begin{equation}\label{resasymplaptr}
\mathcal{R}(x,n) \simeq {\theta \over q^2} ~ x^{-1-\theta} ~ , \qquad x \gg { |\beta|^{1/\theta} \over q^{1/\theta} } ~ .
\end{equation}
The intermediate asymptotic regime corresponds to an interval of  still small values of $y$, but
not too small so that the following inequality $|\beta| y^\theta\gg q$ holds. Then, one can
neglect the term $q$ in the denominator of expression \eqref{resthleson}
to obtain the following intermediate asymptotics
\begin{equation}\label{resolvqeqze}
\mathcal{\tilde{R}}(y,n) \simeq {1 \over |\beta| y^\theta} \quad \mapsto \quad \mathcal{R}(x,n) \simeq {\sin(\pi\theta) \over \pi}~ x^{-1+\theta}~ , \quad  x \ll { |\beta|^{1/\theta} \over q^{1/\theta}}~ .
\end{equation}
The two regimes (\ref{resasymplaptr}) and (\ref{resolvqeqze}) are asymptotics of the inverse Laplace
transform of (\ref{resthleson}) whose explicit expression reads \cite{PSW2005}
\begin{equation}\label{exactoriginthleon}
\mathcal{R}(x,n)= {1 \over q} \left({q \over |\beta|}\right)^{1/\theta} \mathcal{Q} \left(\left({q \over |\beta|}\right)^{1/\theta} x,\theta \right)~ .
\end{equation}
where
\begin{equation}\label{mathcrzerdef}
\mathcal{Q}(x,\theta)= {\sin(\pi\theta) \over \pi} x^{\theta-1} \int_0^\infty { u^\theta e^{-u} du \over u^{2\theta} + x^{2\theta} + 2 x^\theta u^\theta \cos(\pi\theta)} ~ .
\end{equation}
For instance, for $\theta=1/2$,
\begin{equation}\label{rzedefthonehalf}
\mathcal{Q}(x,1/2) =  {1 \over \sqrt{\pi x}} -e^x \text{erfc}(\sqrt{x})~ .
\end{equation}

\subsubsection{Case $1<\theta<2$}

We rewrite (\ref{resolvasymplaim}) as
\begin{equation}\label{resolvasymplaimod}
\mathcal{\tilde{R}}(y,n) \simeq {1 \over q +|\alpha| y -|\beta| y^\theta}  ~ .
\end{equation}

For
\begin{equation}
{1 \over q}~ |\alpha y|\ll 1~,
\end{equation}
then
\begin{equation}\label{thmooneverysmally}
\mathcal{\tilde{R}}(y,n) \simeq {1 \over q} - {|\alpha| \over q^2} y + {|\beta| \over q^2} y^\theta~ ,
\end{equation}
which is the Laplace image of
\begin{equation}\label{asresytoqetgtinf}
\mathcal{R}(x,n) \simeq  {|\beta| \over q^2 \Gamma(-\theta)}~ x^{-\theta-1}
= {\theta \over q^2} ~ x^{-\theta-1}
~ , \qquad x \gg {1 \over q}~ .
\end{equation}
The intermediate asymptotic describes the time interval such that $q \ll |\alpha| y$, leading to
\eqref{resolvasymplaimod}
\begin{equation}\label{resolvasymplaimod}
\mathcal{\tilde{R}}(y,n) \simeq {1 \over |\alpha| y -|\beta| y^\theta}= {1 \over y} \cdot {1 \over |\alpha| - |\beta| y^{\theta-1}} ~ .
\end{equation}
For sufficiently small $|y|$, one gets
\begin{equation}
\mathcal{\tilde{R}}(y,n) \simeq  {1 \over |\alpha|} \cdot {1 \over y} + \left|{\beta \over \alpha^2} \right| \cdot y^{\theta-2} ~ ,
\end{equation}
which is the Laplace image of
\begin{equation}\label{asumpresfromlaptrthmone}
\mathcal{R}(x,n) \simeq {1 \over |\alpha|} + \left|{\beta \over \alpha^2} \right| ~ {1 \over \Gamma(2-\theta)} \cdot x^{-(\theta -1)}~ , \quad  1 \ll x \ll {1 \over q}~
\end{equation}

Expressions (\ref{asumpresfromlaptrthmone}) and (\ref{asresytoqetgtinf}) show that
the resolvent $\mathcal{R}(x,n)$ is a constant plus a weak power law correction $\sim 1/x^{\theta -1}$
for $1 \ll x \ll 1/(1-n)$ which crosses over to $\sim 1/x^{-\theta -1}$ for $x \gg 1/(1-n)$.
A function proportional to the resolvent is plotted in Fig.~4 at the end of the paper, which
shows these different regimes (\ref{asumpresfromlaptrthmone}) and (\ref{asresytoqetgtinf}).

\subsection{Asymptotic of the mean activity}

While the previous subsection provides the expressions of the different regimes of the resolvent,
their derivation using the Laplace transform augmented by different expansions do not provide an
understanding of the derived terms, which we would wish to be based on the underlying
mechanism of cascades of triggering over different generations.
To achieve this goal and remove Paradox 1, we have to explore the asymptotic behavior of the mean activity $\mathcal{A}(x,n)$ for large $x$ values. For this, we study the asymptotics of the corresponding Laplace image $\tilde{\mathcal{A}}(y,n)$ for small $y$ values. Substituting in the series \eqref{Lapimmothres} the asymptotic expression \eqref{genasymplapim} and using the asymptotic relation
\begin{equation}
(1 +\alpha y - \beta y^\theta)^k \simeq e^{k\alpha y - k \beta y^\theta}~ , \qquad |\alpha y - \beta y^\theta| \ll 1~ ,
\end{equation}
we obtain
\begin{equation}\label{geomsermothres}
\tilde{\mathcal{A}}(y,n) \simeq  \sum_{k=1}^\infty n^k \tilde{\psi} (|\beta|^{1/\theta} k^{1/\theta} y;\theta) e^{\alpha k y}~ ,
\end{equation}
where
\begin{equation}\label{psidef}
\tilde{\psi}(y;\theta) = e^{-\text{sign}(\beta) y^\theta} ~ .
\end{equation}
Taking the inverse Laplace transform of the series \eqref{geomsermothres}, we obtain a series representation of the sought mean activity as
\begin{equation}\label{seriesmothres}
\mathcal{A}(x,n) \simeq \sum_{k=1}^\infty \mathcal{A}_k(x,n) ~ ,
\end{equation}
where
\begin{equation}\label{psikththrupsith}
\mathcal{A}_k(x,n) = {n^k \over k^{1/\theta} |\beta|^{1/\theta}} \psi\left({x + \alpha k \over k^{1/\theta} |\beta|^{1/\theta}};\theta \right)~ ,
\end{equation}
and $\psi(x;\theta)$ is a stable distribution, whose Laplace image is given by expression
\eqref{psidef}.

The asymptotic validity of relation \eqref{seriesmothres} for the dependence of $\mathcal{A}(x,n)$ for $x\gg 1$,
describing in particular the case where the kernel is given by expression \eqref{powkern}, transforms
into an exact equality for the mean activity if the kernel of the integral equation \eqref{ainteq} coincides with the stable distribution
\begin{equation}\label{kernelstabdist}
\Phi(t) = {1 \over \varrho} \psi\left({t \over \varrho}; \theta\right)~ ,
\end{equation}
whose Laplace image is given by (\ref{psidef}).

\subsection{Properties of the stable distribution $\psi(x;\theta)$ defined by (\ref{psidef}) and (\ref{kernelstabdist})}

Formulas \eqref{seriesmothres} and \eqref{psikththrupsith} imply that a better understanding of the asymptotic shape of the mean activity $\mathcal{A}(x,n)$ is dependent on a detailed knowledge of the properties of the stable distribution $\psi(x;\theta)$.
This subsection is devoted to this question.

There are many integral representations of the stable distribution $\psi(x;\theta)$. In particular, one can show that
\begin{equation}\label{stableintrepr}
\begin{array}{c}\displaystyle
\psi(x;\theta) =
{1 \over \pi} \int_0^\infty \exp\left(-\big|\cos\left({\pi \theta \over 2}\right)\big| u^\theta \right) \times
\\[5mm] \displaystyle
\cos\left(u x+ u^\theta \sin\left({\pi \theta \over 2}\right) \text{sign}(\theta-1)\right) du~ , \qquad 0<\theta<2~ , \quad \theta\neq 1~ .
\end{array}
\end{equation}

Explicit analytic expressions of the stable distribution $\psi(x;\theta)$ exist for
some specific values of the parameter $\theta$.  For illustrative purposes, we will use below two
such stable distributions. The first one is the famous Levy stable law
\begin{equation}\label{levystabdist}
\psi(x;1/2) = {1 \over 2 x \sqrt{\pi x}} \exp\left(-{1 \over 4 x} \right)~ ,~~~~~(\theta =1/2)~,
\end{equation}
and the other is
\begin{equation}\label{stabthreetwo}
\begin{array}{c}
\psi(x;3/2) =
\\[4mm]\displaystyle
{1 \over \pi \sqrt{3}} \left[\Gamma\left({2 \over 3} \right) \mathstrut_1\! F_1\left({5 \over 6}, {2 \over 3}, {4 x^3 \over 27}\right) - x \Gamma\left({4 \over 3}\right) \mathstrut_1\! F_1\left({7 \over 6}, {4 \over 3}, {4 x^3 \over 27}\right) \right]~ ,~~~~~(\theta=3/2)~,
\end{array}
\end{equation}
where $ \mathstrut_1\! F_1\left(a, b, c \right)$ denote a confluent hypergeometric function of the first kind.

All stable laws possess in common \emph{long} and \emph{short} tails.
By definition, the long tail is their power law behavior at $x\to\infty$:
\begin{equation}\label{powerlawstabdist}
\psi(x;\theta) \simeq {x^{-\theta -1 } \over \Big|\Gamma(-\theta)\Big|}~ , \qquad x\to\infty ~ .
\end{equation}
These short tail of stable distributions consists in a very fast decay of $\psi(x;\theta)$ to zero as $x\to 0$ for $0<\theta<1$ corresponding to an essential singularity at $x=0$, and in their super-exponentially fast decay as $x\to -\infty$ for $1<\theta<2$. The following asymptotic formula is true \cite{Uchaikin}
\begin{equation}\label{asympshort}
\begin{array}{c} \displaystyle
\psi(x;\theta) \simeq {1 \over \sqrt{2\pi \theta |\theta-1|}} \left({|x| \over \theta}\right)^{{2-\theta \over 2\theta -2}} \exp\left(-|\theta-1| \left({|x| \over \theta}\right)^{{\theta \over \theta-1}} \right)~ ,
\\[8mm]\displaystyle
\begin{cases}
x\to 0_+~ , & \quad \text{if} \quad 0<\theta<1 ~ , \\[2mm]
x\to -\infty ~ , & \quad \text{if} \quad 1<\theta<2 ~ .
\end{cases}
\end{array}
\end{equation}
Curiously, for $\theta=1/2$, this asymptotic formula coincides with the Levy stable law \eqref{levystabdist}.  According to the asymptotic relation \eqref{asympshort}, the stable distribution $\psi(x;3/2)$ decays at $x\to-\infty$ according to
\begin{equation}\label{asympthreetwo}
\psi(x;3/2) \simeq \sqrt{4 |x| \over 9 \pi } \exp\left(- {4 \over 27} |x|^3 \right) ~ , \qquad x \to -\infty ~ .
\end{equation}
In practice, the asymptotic expression such as (\ref{asympthreetwo}) can be verified to be extremely
accurate already for $x<-2$.

\section{Solution of Paradox 1 for  $0<\theta<1$}

The asymptotic behavior  for $x\gg 1$ of the mean activity $\mathcal{A}(x,n)$ given by \eqref{seriesmothres} is qualitatively different for $\theta\in(0,1)$ and for $\theta\in(1,2)$. In this section, we focus on the former case $\theta\in(0,1)$.

\subsection{Integral approximation of the mean activity $\mathcal{A}(x,n)$}

As summarized in the statement of Paradox 1, for $\theta\in(0,1)$ and close to criticality ($n\lesssim 1$), the mean activity $\mathcal{A}(x,n)$ exhibits a double power law behavior, with the coexistence of the power law asymptotic
\begin{equation}\label{powoneminth}
\mathcal{A}(x,n) \sim x^{-1-\theta} ~ , \qquad x\gg 1~ .
\end{equation}
for very large $x$ values, and of an intermediate asymptotic regime for smaller $x$ (but still remaining large)
\begin{equation}\label{powoneplth}
\mathcal{A}(x,n) \sim x^{-1+\theta} ~ , \qquad  {\rm intermediate ~asympotics}~ .
\end{equation}
The goal of this subsection is to show that the intermediate power asymptotic \eqref{powoneplth} results from the fast decay of the short tail part of the stable distribution $\psi(x;\theta)$, which controls the mean activity
$\mathcal{A}_k(x,n)$ defined by \eqref{psikththrupsith} of the activity resulting from the $k$-th generation.
In contrast, we will show that the  power asymptotic (\ref{powoneminth}) is due to the
corresponding power asymptotic of the bare kernel $\Phi(t) \sim t^{-\theta-1}$.

The first step consists in noting that, for $0<\theta<1$, the shift in expression \eqref{psikththrupsith} for $\mathcal{A}_k(x,n)$
can be written as
\begin{equation}\label{shiftpar}
{\alpha k \over |\beta|^{1/\theta} k^{1/\theta}} \sim k^{{\theta-1 \over \theta}}~.
\end{equation}
It thus tends to zero for $k\to\infty$, so that one may neglect it as it will not impact the asymptotic law of the mean activity $\mathcal{A}(x,n)$ for large $x$'s. In other words, for $0<\theta<1$, one may without essential error replace the $k$-th generation mean activity $\mathcal{A}(x,n)$ by
\begin{equation}\label{kthgennotshift}
\mathcal{A}_k(x,n) = {n^k \over k^{1/\theta} |\beta|^{1/\theta}} \psi\left({x\over k^{1/\theta} |\beta|^{1/\theta}};\theta \right)~ .
\end{equation}

The next step is to notice that, if $x$ large enough, then $\mathcal{A}_k(x,n)$ becomes a sufficiently smooth function of the argument $k$ and one may, without essential error, replace the series \eqref{seriesmothres} by the integral
\begin{equation}\label{abodyint}
\mathcal{A}(x,n) \simeq \int\limits_1^\infty \mathcal{A}_k(x,n)dk \simeq \int\limits_1^\infty {e^{-\gamma k} \over |\beta|^{1/\theta} k^{1/\theta}} \psi\left( {x \over |\beta|^{1/\theta} k^{1/\theta}};\theta \right) dk~ ,
\end{equation}
in which we have defined
\begin{equation}
\gamma = \ln\left({1 \over n}\right)~ .
\label{hythywa}
\end{equation}
We suppose everywhere below that $|\gamma|\ll 1$ ($n$ is close to its critical value $1$), so that the exponential function $e^{-\gamma k}$ in the integral \eqref{abodyint} is also a smooth function of $k$.

Using the change of variable
\begin{equation}
u = {x \over |\beta|^{1/\theta} k^{1/\theta}} \quad \iff \quad k = {1 \over |\beta|} \left({x \over u}\right)^\theta~ , \quad dk = - {\theta \over |\beta|} \left({x \over u}\right)^\theta {du \over u}~ ,
\end{equation}
the integral \eqref{abodyint} becomes
\begin{equation}\label{axnintu}
\mathcal{A}(x,n) \simeq {1 \over x^{1-\theta}} {\theta \over |\beta|} \int_0^{u(x)} \exp\left( - {\gamma \over |\beta|} \left({x \over u} \right)^\theta \right)
\psi(u;\theta) {du \over u^\theta}~ .
\end{equation}
where
\begin{equation}
u(x) = {x \over |\beta|^{1/\theta}}~ .
\end{equation}
As $u(x)\gg 1$ for $x \gg 1$, one may replace without essential error the upper limit in the integral \eqref{axnintu} by infinity:
\begin{equation}\label{axnintuinf}
\mathcal{A}(x,n) \simeq {1 \over x^{1-\theta}} {\theta \over |\beta|} \int_0^\infty \exp\left( - {\gamma \over |\beta|} \left({x \over u} \right)^\theta \right)
\psi(u;\theta) {du \over u^\theta}~ .
\end{equation}
For $\gamma=0$ ($n=1$), we obtain the mean activity given by \eqref{axnintuinf} as the power law \eqref{powoneplth} corresponding to the intermediate asymptotics:
\begin{equation}\label{intpowthlesone}
\mathcal{A}_\text{int}(x,n) \simeq {C(\theta) \over x^{1-\theta}} {\theta\over|\beta|}~,
\end{equation}
where the index `int' refers to ``intermediate asymptotics'' and
\begin{equation}
C(\theta) = \int_0^\infty \psi(u;\theta) {du \over u^\theta} = {1 \over \Gamma(1+\theta)}~ .
\end{equation}
It is reasonable to choose $\beta= \Gamma(1-\theta)$ defined in (\ref{albesigns}) and using the well-known identity
$\Gamma(1-\theta) \Gamma(1+\theta) \equiv \pi \theta \csc(\pi\theta)$, we obtain
\begin{equation}\label{aintermasympsosec}
\mathcal{A}_\text{int}(x,n) \simeq {\sin(\pi\theta) \over \pi}  \cdot x^{-1+\theta} ~ .
\end{equation}
For  $\gamma=0$ ($n=1$), the intermediate asymptotic (\ref{aintermasympsosec}) invades the whole
large $x$ regime, so that its determination is clearer.

The fact that expression (\ref{aintermasympsosec})
holds only for an intermediate range of $x$'s values for $\gamma>0$ ($n<1$) is now determined from
the following derivation.
Let $\xi(\theta)$ be such that, for if $x>\xi(\theta)$, the stable distribution $\psi(x;\theta)$ is not different from its long tail \eqref{powerlawstabdist} within a specified error margin. For example, for the
Levy stable distribution $\psi(x;1/2)$ given by \eqref{levystabdist}, $\xi(\theta)$ can be taken equal to $3$
when considering an error margin of less than $1\%$.

From expression \eqref{axnintuinf}, one can see that, if the following condition holds,
\begin{equation}\label{ineqforlastpowas}
{\gamma \over |\beta|} \left({x \over \xi(\theta)}\right)^\theta \gtrsim 1 \quad \Rightarrow \quad x \gtrsim \xi(\theta) \left({|\beta| \over \gamma} \right)^{1/\theta}
\end{equation}
then, without essential error, the stable distribution $\psi(x;\theta)$ can be replaced by its long tail in the integral \eqref{axnintuinf}.
This leads to the following approximate relation
\begin{equation}\label{ainteventasymp}
\mathcal{A}(x,n) \simeq {1 \over x^{1-\theta}} {\theta \over |\beta| |\Gamma(-\theta)|} \int_0^\infty \exp\left( - {\gamma \over |\beta|} \left({x \over u} \right)^\theta \right) {du \over u^{2\theta+1}}~ .
\end{equation}
Changing the integration variable to $z = u/x$, we obtain
\begin{equation}\label{ainteventasympinf}
\mathcal{A}(x,n) \simeq {1 \over x^{1+\theta}} {\theta D(\theta) \over |\beta| |\Gamma(-\theta)|}~ ,
\end{equation}
where
\begin{equation}
D(\theta) = \int_0^\infty \exp\left( -{\gamma \over |\beta| z^\theta} \right) {dz \over z^{2\theta+1}} = {\beta^2 \over \theta \gamma^2}~ .
\end{equation}

\subsection{Dissecting the two key contributions to solve Paradox 1}
\label{theyt4ns}

The asymptotic analysis of the mean activity $\mathcal{A}(x,n)$ presented in the previous subsection, while sufficiently
rigorous, does not provide an intuitive understanding of the two intermediate and asymptotic regimes and of the solution
of Paradox 1. In the present subsection, we provide a cruder but more transparent analysis, which reveals
the hidden springs of the crossover from the intermediate asymptotic power law \eqref{powoneplth} to the full asymptotic power law\eqref{powoneminth}. The next section will provide a different approach which
illuminates even further the mechanism of the transition from the bare kernel time decay to the resolvent time dependence.

For this, we replace in expression \eqref{psikththrupsith} the stable distribution $\psi(x;\theta)$ by its ``geometrical power law skeleton''
\begin{equation}\label{phithelessoneskel}
\psi_0(x;\theta) =
\begin{cases}\displaystyle
{x^{-1-\theta} \over |\Gamma(-\theta)|}~ , & x>\xi(\theta)~ , \\[3mm]
\displaystyle
0 ~ , & x< \xi(\theta)~ .
\end{cases}
\end{equation}
Replacing in \eqref{psikththrupsith} $\psi(x;\theta)$ by $\psi_0(x;\theta)$ (and neglecting the shift parameter \eqref{shiftpar}), we obtain
\begin{equation}\label{aktasympk}
\mathcal{A}_k^0(x,n) \simeq x^{-1-\theta} n^k {|\beta| k\ \over |\Gamma(-\theta)|} \bs{1}(x-|\beta|^{1/\theta} k^{1/\theta} \xi(\theta))~ ,
\end{equation}
where $\bs{1}(z)$ is the unit step function. This term $\bs{1}(x-|\beta|^{1/\theta} k^{1/\theta} \xi(\theta))$ is important, as
it accounts semi-quantitatively for the fast decaying short tail of the stable distribution $\psi(x;\theta)$, which
has the role of effectively truncating the series \eqref{seriesmothres} at large $k$'s.

For concreteness, let us consider the particular kernel $\Phi(t)$ defined by \eqref{powkern}, for which $\beta=\Gamma(1-\theta)$.
We can then rewrite relation \eqref{aktasympk} in the more transparent form
\begin{equation}\label{aktasympev}
\mathcal{A}_k^0(x,n) = \theta ~ x^{-1-\theta} ~ k ~ n^k ~\bs{1}(k(x,\theta) - k) ~ ,
\end{equation}
where
\begin{equation}\label{kxth}
k(x,\theta) = {x^\theta \over \xi^\theta(\theta) \beta} = {x^\theta \over \xi^\theta(\theta) \Gamma(1-\theta)} ~ .
\end{equation}
Choosing for simplicity $\xi^\theta(\theta)= 1/|\beta|$, we obtain
\begin{equation}
\label{rgwqeqa}
k(x,\theta) = x^\theta~ .
\end{equation}
Substituting expression \eqref{aktasympev} into the series \eqref{psikththrupsith}, we obtain the mean activity
estimated in this geometrical skeleton approximation, denoted as $\mathcal{A}^0(x,n)$, as
\begin{equation}\label{atsumasymp}
\mathcal{A}^0(x,n) \simeq \theta ~ x^{-1-\theta} ~ \mathcal{S}(k(x,\theta),n) = \theta~ x^{-1-\theta} \cdot \mathcal{S}(x^\theta,n)~ ,
\end{equation}
where
\begin{equation}\label{mathsdef}
\mathcal{S}(\kappa,n) = \sum_{k=1}^{\kappa} k\cdot n^k ~ .
\end{equation}

Expression (\ref{atsumasymp}) with (\ref{mathsdef}) allows us to pinpoint the origin of the
slow decay $\sim 1/t^{1-\theta}$ of the resolvent in the intermediate asymptotic regime or for $n=1$
as due to the fight between the fast decay  $\sim 1/t^{1+\theta}$ of the bare kernel and
the growth $\sim t^{2\theta}$ of the contributions to the activity at time $t$ of all generations set in motion up to time $t$.
This later growth is controlled by the short tail of the stable distribution $\psi(x;\theta)$ corresponding
to the truncation on the above geometrical skeleton approximation (\ref{phithelessoneskel}).
The main contributions to $\mathcal{A}(x,n)$ are provided by the first $k(x,\theta)$ summands,  because the
mean activities $\mathcal{A}_k(x,n)$ of the highest order generations, for which $k\gtrsim k(x,\theta)$, are, for given $x=t/\varrho$, not yet large enough to influence significantly the total activity level $A(t)$. Roughly speaking, the larger the order $k$ of a generation, the later its contribution $\mathcal{A}_k(x,n)$ is felt.

For $n=1$, the sum \eqref{mathsdef} reduces to
\begin{equation}\label{svalcritcase}
\mathcal{S}(\kappa,1) = \sum_{k=1}^{\kappa} k = {1 \over 2} \kappa (\kappa+1) \simeq {1 \over 2} \kappa^2~ .
\end{equation}
Using (\ref{rgwqeqa}), we obtain
\begin{equation}\label{sumsexplexpr}
\mathcal{S}(k(x,\theta),1) \simeq {1 \over 2} x^{2\theta}~ .
\end{equation}
Substituting this expression (\ref{sumsexplexpr}) into \eqref{atsumasymp}, we finally obtain
promised power law \eqref{powoneplth}
\begin{equation}\label{atsumasympo}
\mathcal{A}^0(x,n) \simeq {\theta \over 2} \cdot x^{-1-\theta} \cdot x^{2\theta} \simeq {\theta \over 2} \cdot x^{-1+\theta}~ .
\end{equation}

This last equation (\ref{atsumasympo}) illuminates the origin of Paradox 1: the larger $\theta$ is, the faster
the decay of the bare kernel $\sim 1/t^{1+\theta}$, but the larger the number $k(t,\theta) \sim t^\theta$ of generations
which are activated up to time $t$ and the greater their combined contribution $\sim [k(t,\theta)]^2 \sim t^{2\theta}$
to the overall activity at time $t$, so that, all being taken into account, the response function
actually develops a longer memory
$\sim  t^{2\theta} \times 1/t^{1+\theta} = 1/t^{1-\theta}$. Paradox 1 can
thus be seen as a result of an ``anomalously'' slow triggering of successive
generations associated with the infinite average waiting time between triggered events. Indeed,
the average waiting time between two events, as described by the bare kernel, is defined by
\begin{equation}
\langle t \rangle \sim {\rm lim}_{T \to +\infty} \int^T {t \over t^{1+\theta}} dt~,
\label{jhigkfmpwa}
\end{equation}
which is diverging as the upper bound $T$ of the integral goes to infinity, for $\theta \leq 1$.
This divergence is a standard diagnostic of the existence of an anomalous
trapping time regime \cite{BouchaudGeorges,Sornettecrit06}, leading to anomalous scaling laws.
In the present case, the ``anomalous'' scaling law is the ``renormalization'' of the bare
kernel time decay $\sim 1/t^{1+\theta}$ into the resolvent time decay $\sim 1/t^{1-\theta}$.
The next subsection \ref{intuitionparadox} re-derives this result from scratch by using a completely
intuitive and straightforward reasoning, exemplifying that the root of Paradox 1 indeed
lies on the diverging mean waiting time between triggered events and its associated
anomalous diffusion.

But before doing so, we exploit the present analysis to describe the
subcritical case $n\lesssim 1$. For arbitrary $n$, the sum \eqref{mathsdef} is equal to
\begin{equation}\label{snthryq}
\mathcal{S}(\kappa,n) = {1 -q \over q^2} \left[ 1-(1-q)^\kappa (1+\kappa \cdot q)\right]~ , \qquad q = 1-n~ .
\end{equation}
For small enough $q$, such that $\kappa q \ll 1$, expression \eqref{snthryq} has the same square
asymptotic \eqref{svalcritcase} as for $n=1$. In contrast, for $\kappa q\gg 1$,  $\mathcal{S}(\kappa,n)$ tends to a constant limit
$\mathcal{S}(\kappa,n) \to {1 \over q^2}$. This implies that, when $k(x,\theta) q \gg 1$,
the mean activity $\mathcal{A}^0(x,n)$ obeys the same power law as the bare kernel,
\begin{equation}
\mathcal{A}^0(x,n) \simeq  { \theta \over q^2} \cdot x^{-1-\theta}~ ,
\end{equation}
as seen from \eqref{atsumasymp}.
The function $\mathcal{S}(\kappa,n)$ defined by \eqref{snthryq} is plotted in Fig.~1 for different $q=1-n$ values. It demonstrates the crossover from $\mathcal{S}(\kappa,n)\sim \kappa^2$ for  $\kappa q \ll 1$ to $\mathcal{S}(\kappa,n) \simeq \text{const}$ for $\kappa q\gg 1$. This crossover governs the crossover of the resolvent from $1/t^{1-\theta}$ for $t \lesssim 1/(1-n)^{1/\theta}$  to  $1/t^{1+\theta}$  for $t \gtrsim 1/(1-n)^{1/\theta}$.

\begin{quote}
\centerline{
\includegraphics[width=11cm]{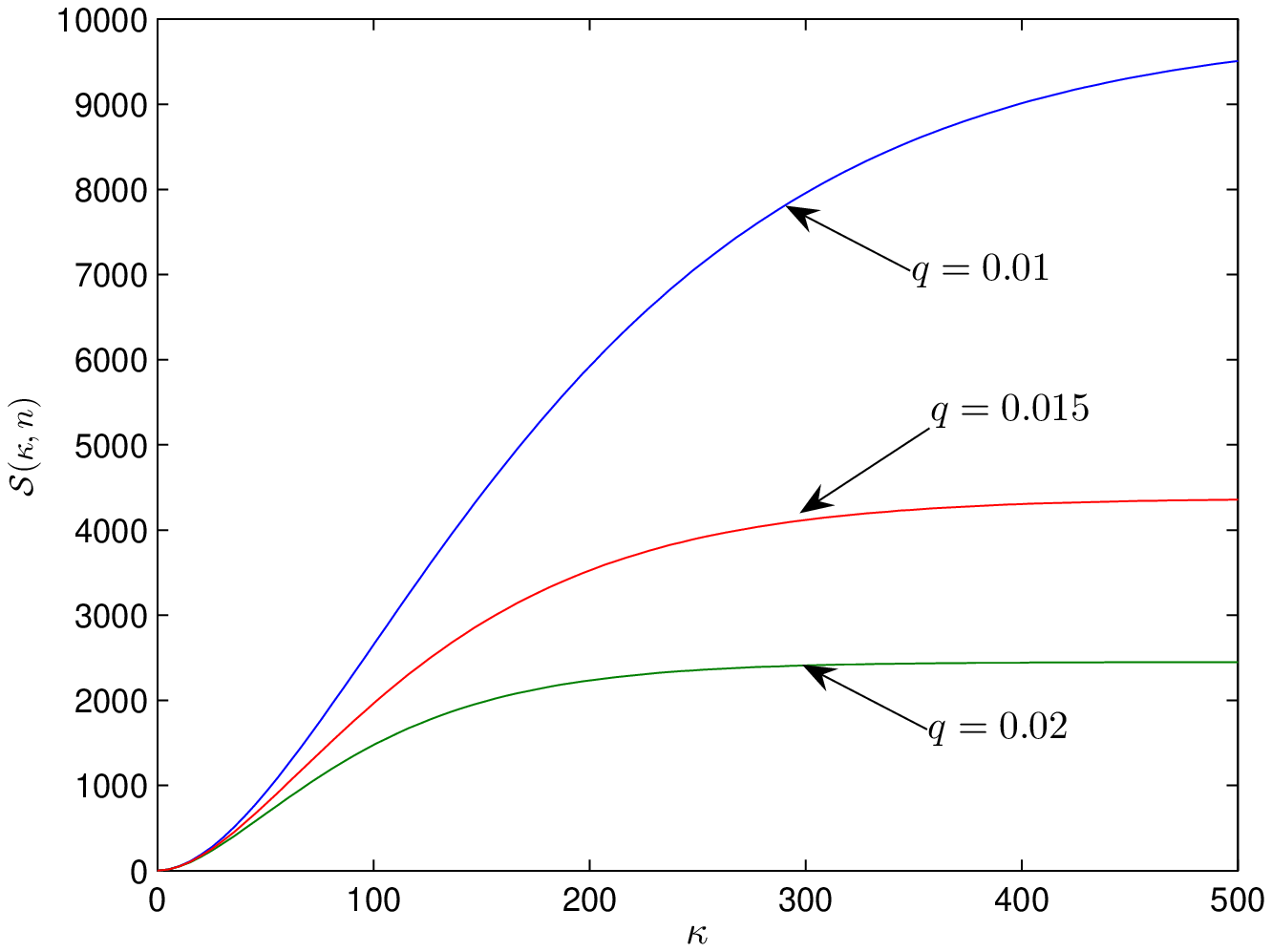}}
{\bf Fig.~1:} \small{Plot of function $\mathcal{S}(\kappa,n)$ defined by \eqref{snthryq} for $q \equiv 1-n=0.01; 0.015; 0.02$, illustrating the cross-over between the power quadratic behavior (\ref{svalcritcase}) for  $\kappa q \ll 1$ to a constant
for $\kappa q\gg 1$.}
\end{quote}

\subsection{Intuitive derivation explaining Paradox 1 based on scaling arguments for long waiting times} \label{intuitionparadox}

Let us now re-derive all the key results of the previous sub-section by a completely different and intuitive route.
Our approach is based on the conceptual view of the total activity $A(t)$ at a given time $t$ as the superposition
of the activities $\mathcal{A}_k(t)$ coming from all possible generations $k=1, 2,...$ that are significant at this time $t$.

Consider a first burst of activity starting at time $0$, constituting the event of zero-th generation.
This initial event may lead to an event of first
generation at a latter time $t_1$, which itself may trigger an event of second generation at time $t_2$, and so on.
We assume that these events constitute the starting time for each successive generation to contribute
significantly to the overall activity. For a given time $t$ of observation of the activity,
our problem is to determine the typical time $t_k(t)$ of occurrence of the $k$-th generation
and its corresponding contribution $\mathcal{A}_k(t)$.

For the first question, we use the interpretation that the bare kernel $\Phi(t)$ is nothing but
the probability density function (pdf) of the waiting time $t$
from a burst of activity and its first triggered event.
Let us suppose that $K(t)$ generations have been triggered over the total time interval $t$.
Time consistency imposes that
\begin{equation}
\label{tgwtrgrwaaftr}
t_1 + t_2 +... + t_{K(t)} = t~.
\end{equation}
Let us call $t_{\rm max}(t)$ the largest waiting time among the $K(t)$ values $ t_1, t_2, ... ,t_{K(t)}$.
Since the probability that a waiting time between two successive generations
is equal to or larger than $t_{\rm max}(t)$ is of the order of $\varrho^\theta\int_{t_{\rm max}(t)}^{+\infty} dt / t^{1+\theta}$, by consistency, one must have
\begin{equation}
K(t) \times \varrho^\theta\int_{t_{\rm max}(t)}^{+\infty} dt / t^{1+\theta} \sim 1~.
\label{thgtrgwrq}
\end{equation}
Expression (\ref{thgtrgwrq}) just states that there is typically just one waiting time of the order of the maximum
waiting time $t_{\rm max}(t)$ among the $K(t)$ waiting times between the successive generations. The solution of (\ref{thgtrgwrq})  is
\begin{equation}
t_{\rm max}(t) \sim \varrho [K(t)]^{1/\theta}~.
\end{equation}
The dependence $K(t)$ as a function of $t$ is then obtained by estimating the l.h.s of (\ref{tgwtrgrwaaftr}) as
\begin{equation}
t_1 + t_2 +... + t_{K(t)} = K(t) \times \langle t \rangle_t  \sim K(t) \times \varrho^\theta\int_{0}^{t_{\rm max}(t)} dt~ t / t^{1+\theta} \sim \varrho [K(t)]^{1/\theta}~.
\label{thwtrbw}
\end{equation}
We have used the fact that the average waiting time $\langle t \rangle_t $ between successive generations
has to be estimated by the standard  sum of all possible $t$'s weighted by their corresponding
probability, but with an upper bound since no waiting times larger
than $t_{\rm max}(t)$ are sampled in the finite set of $K(t)$ realizations.  This trick is standard to tame
the infinities of the unconditional average waiting time $\langle t \rangle$ defined by (\ref{jhigkfmpwa})
in the limit $T \to +\infty$, found for $\theta <1$, leading
to anomalous diffusion and other abnormal scaling effects \cite{BouchaudGeorges,Sornettecrit06}.
Then, by (\ref{tgwtrgrwaaftr}), we obtain $[K(t)]^{1/\theta} \sim t/\varrho$ and thus
\begin{equation}
K(t) \sim (t/\varrho)^\theta~,
\label{htyhtne}
\end{equation}
which retrieves (\ref{rgwqeqa}) obtained in the previous sub-section.

The contribution $\mathcal{A}_k(t)$ of the $k$-th active generation at time $t$ has two
important terms. The first one is the probability $n^k$ that $k$ generations have occurred.
The second one is based on
the concept that each activated generation contributes proportionally to the bare kernel $\sim 1/t^{1+\theta}$
but with a characteristic time scale $\rho_k$ equal to the only existing time scale associated with
the generation $k$, namely its waiting time $t_k$: $\rho_k \sim t_k$ until its happenance.  In complete analogy with the shape of $\Phi(t)$  given by (\ref{powkern0}), this leads finally to
\begin{equation}
\mathcal{A}_{k} \sim n^k~ {\theta [t_k(t)]^\theta \over (t+t_k(t))^{1+\theta}}~.
\label{thb3sd}
\end{equation}
Ordering the indices by increasing values of the waiting times $t_k(t)$, expression (\ref{htyhtne}) implies
that $t_k(t) \sim \varrho k^{1/\theta}$, and thus
\begin{equation}
\mathcal{A}_{k} \sim n^k~{\theta k \over t^{1+\theta}}~,~~~~{\rm for}~t \gg t_k(t)~.
\label{h3iqpooadkv}
\end{equation}
The total activity is thus
\begin{equation}
\mathcal{A}(t) = \sum_{k=1}^{K(t)} \mathcal{A}_k(t) \sim {\theta \over t^{1+\theta}} \times  \sum_{k=1}^{K(t)} k n^k~,
\label{hjuwosk}
\end{equation}
which recovers (\ref{atsumasymp}) with (\ref{mathsdef}).

\subsection{Validation with the exactly solvable case $\theta =1/2$}

It is always useful to check the validity of asymptotic relations when exact results are available.
Here, we compare the above asymptotic relations of subsection \ref{theyt4ns} for
the mean activity $\mathcal{A}(x,n)$ with the series \eqref{seriesmothres}, \eqref{psikththrupsith} for $\theta=1/2$. In this case, the series \eqref{seriesmothres} takes the form
\begin{equation}\label{atthethalf}
\mathcal{A}(x,n) = \sum_{k=1}^\infty {1 \over \pi k^2} \psi\left({x+2 k \over \pi k^2};{1 \over 2} \right) ~ .
\end{equation}
while its geometrical skeleton is equal to
\begin{equation}\label{atthethalfsk}
\mathcal{A}^0(x,n) = {1 \over 2} x^{-3/2} \mathcal{S}(\sqrt{x},n)~ .
\end{equation}
A more accurate approximation of the series \eqref{atthethalf} than that given by integral \eqref{axnintuinf}
in the case $\theta=1/2$ is equal to
\begin{equation}\label{aintthonehalf}
\mathcal{A}_\text{int}(x,n) = {1 \over \pi \sqrt{x}} \left[1 -\gamma \sqrt{x} \exp\left({x \gamma^2 \over \pi} \right) \text{erfc} \left({\gamma\sqrt{{x \over \pi}}}\right) \right]~ ,
\end{equation}
where $\gamma$ is defined by (\ref{hythywa}).
For $\theta=1/2$, expressions \eqref{intpowthlesone} and \eqref{ainteventasympinf} become
\begin{equation}\label{aintbothasympt}
\mathcal{A}_\text{int}(x,n) \simeq
\begin{cases}\displaystyle
{1 \over \pi \sqrt{x}}~ , & \gamma = 0 ~ , \\[4mm] \displaystyle
{1 \over 2 x \sqrt{x} \gamma^2}~ , & \displaystyle
x\gtrsim {3 \pi \over \gamma^2}~ .
\end{cases}
\end{equation}
Fig.~2 compares the mean activity $\mathcal{A}(x,n)$ given by \eqref{atthethalf}, its geometric skeleton $\mathcal{A}^0(x,n)$ obtained from \eqref{atthethalfsk} and its integral approximation $\mathcal{A}_\text{int}(x,n)$ given by \eqref{aintthonehalf}. One can check that they are practically undistinguishable for any $x\gtrsim 10$.
One can also observe the two power law regimes $\mathcal{A}\sim x^{-1\pm\theta}$ and their cross-over.

\begin{quote}
\centerline{
\includegraphics[width=11cm]{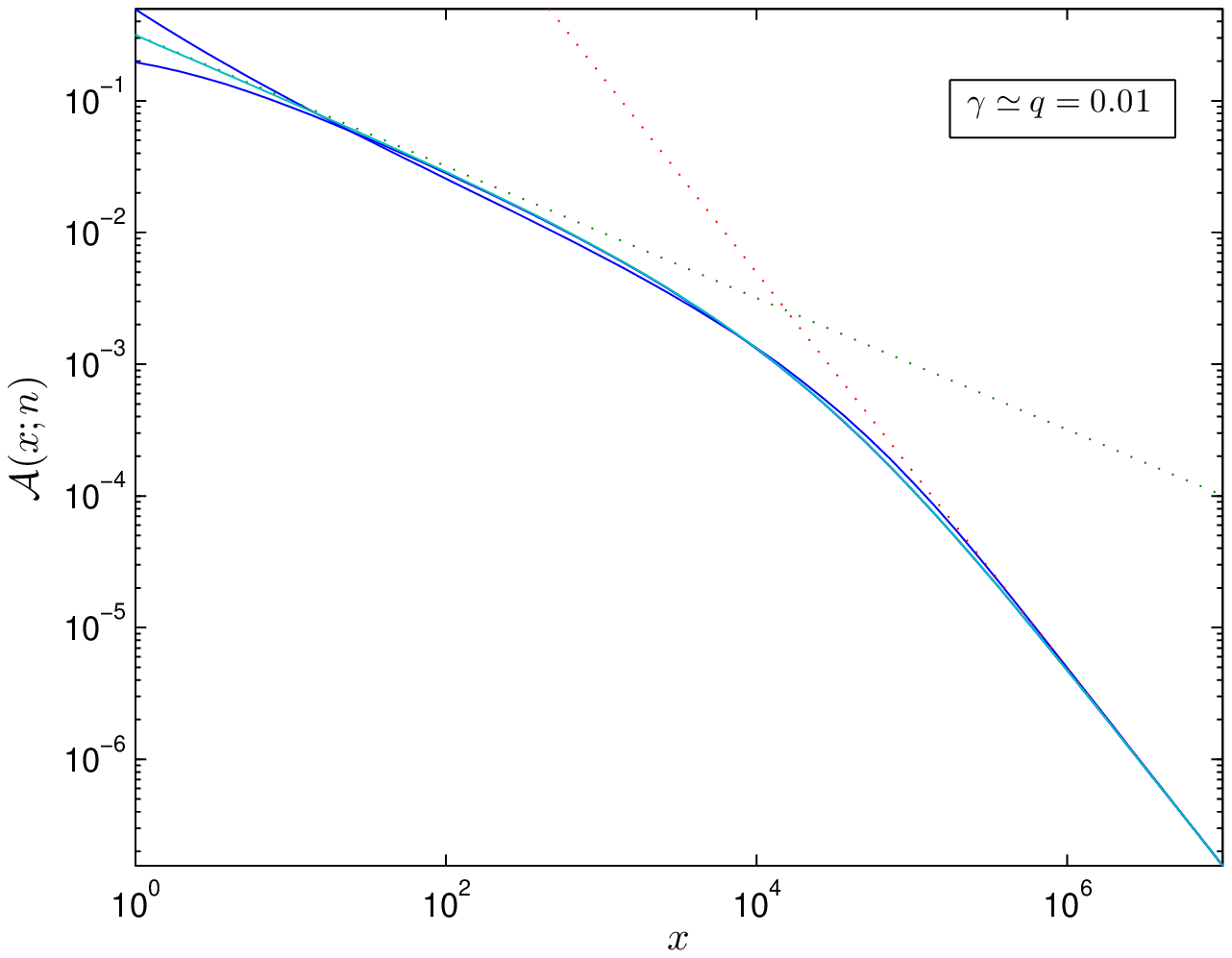}}
{\bf Fig.~2:} \small{Plots of the mean activity $\mathcal{A}(x,n)$ given by \eqref{atthethalf}, its geometric skeleton $\mathcal{A}^0(x,n)$ obtained from \eqref{atthethalfsk} and its integral approximation $\mathcal{A}_\text{int}(x,n)$ given by \eqref{aintthonehalf}, for $\theta=1/2$ and $\gamma\simeq q=1-n=0.01$. The dotted straight lines show the limiting power asymptotics~\eqref{aintbothasympt} corresponding to $\mathcal{A} \sim x^{-1\pm \theta}$.}
\end{quote}

\section{Problem 2: power law exponent for the resolvent for  $1<\theta$}

\subsection{Integral representation}

It is convenient to re-express equations \eqref{seriesmothres} and \eqref{psikththrupsith} in a form
more adapted to the case $\theta>1$:
\begin{equation}\label{atasympstabthmone}
\mathcal{A}(x,n) = \sum_{k=1}^\infty {n^k \over |\beta|^{1/\theta} k^{1/\theta}} \psi\left( {x-|\alpha| k \over |\beta|^{1/\theta} k^{1/\theta}};\theta \right) ~ .
\end{equation}
For large $x \gg |\alpha|$, one may, without significant error, replace the series in  \eqref{atasympstabthmone} by the integral
\begin{equation}\label{athmooneint}
\mathcal{A}_\text{int}(x,n) \simeq
\int_{1}^\infty e^{-\gamma k}~\mathcal{A}_k(x;\theta) dk~ ,
\end{equation}
where
\begin{equation}\label{summandfuncdef}
\mathcal{A}_k(x;\theta) = {1 \over |\beta|^{1/\theta} k^{1/\theta}} \psi\left( {x-|\alpha| k \over |\beta|^{1/\theta} k^{1/\theta}};\theta \right) .
\end{equation}
The ``body'' of the stable distribution $\psi(u;\theta)$ is concentrated near $u=0$, which means that the  ``body'' of $\mathcal{A}_k(x;\theta)$ taken as a function of $k$ is concentrated in the vicinity of
\begin{equation}\label{zeropointd}
k_* = { x \over |\alpha|}~ .
\end{equation}
As a consequence, the value of the integral \eqref{athmooneint} is qualitatively different depending on the value of the parameter
\begin{equation}\label{epsdef}
\varepsilon = \gamma k_* = {\gamma x \over |\alpha|}~ .
\end{equation}

\subsection{Early time asymptotic}

Let us first consider the regime $\varepsilon \ll 1$. In this case, one may, without significant error, put
$\gamma=0$ in \eqref{athmooneint} to obtain the following approximate relation
\begin{equation}\label{athmooneintgamz}
\mathcal{A}_\text{int}(x,n) \simeq
\int_{1}^\infty \mathcal{A}_k(x;\theta) dk~ .
\end{equation}
Given the definition (\ref{hythywa}), putting $\gamma=0$ is equivalent to neglecting the
difference between the branching ratio $n$ and its critical value $1$. In other words, for times sufficiently
short such that $\varepsilon$ defined by (\ref{epsdef}) is small, the mean activity
is faithfully described as if the system was in the critical regime $n=1$.
Taking into account that the effective width of the function $\mathcal{A}_k(x;\theta)$ of the argument $k$, defined
as the domain in which $\mathcal{A}_k(x;\theta)$ is significantly different from zero,  is much smaller than $k_*$, one can replace $k$ by the constant $k_*$ given by \eqref{zeropointd} in the denominators of the r.h.s. of expression \eqref{summandfuncdef}. Then, using the change of integration variable
$k \mapsto  u = {x-|\alpha| k \over |\beta|^{1/\theta} k_*^{1/\theta}}$, we
rewrite the integral \eqref{athmooneint} in the approximate form
\begin{equation}\label{aintconst}
\mathcal{A}_\text{int}(x,n) \simeq {1 \over |\alpha|} \int_{-\infty}^\infty \psi(u;\theta) du = {1 \over |\alpha|} \qquad (\varepsilon \ll 1)~ ,
\end{equation}
as  result of the normalization condition of the stable distribution $\psi(x;\theta)$. In sum, we have
\begin{equation}\label{meanrasympmxleonepl}
\mathcal{A}(x,n) \simeq {1 \over |\alpha|} = \theta -1~ , \qquad {\gamma x \over|\alpha|}\ll 1  ~~ \Longrightarrow  ~~ x \ll {1 \over (\theta-1)(1-n)}~ ,
\end{equation}
where we have used definition (\ref{albesigns}) for $\alpha$ and (\ref{hythywa}) for $\gamma$, assuming that $n$ is close to $1$ so that $\gamma \approx 1-n$.
The result (\ref{meanrasympmxleonepl}) expresses that, for $n\simeq1$, there is \textbf{plateau} in the mean activity $\mathcal{A}(x,n)$
as a function time, for early time $x \ll {1 \over (\theta-1)(1-n)}$.  As $n$ moves closer and closer to $1$, the
regime where $\mathcal{A}(x,n) \simeq \theta -1$ extends to longer and longer times.
We note that this constancy of the resolvent at criticality $n=1$ is well-known for 
an exponentially decaying kernel function $\Phi(t) \sim \exp[-r t]$, corresponding to the resolvent 
$R(t) \sim \exp[-r(1-n)t]$. The novel behavior found here is the existence of a non-trivial cross-over
to the power law $\sim 1/t^{1+\theta}$, as explained in the next subsection \ref{tyhjrgbgrv}.

This result (\ref{meanrasympmxleonepl}) can be recovered simply by using the arguments of subsection
\ref{intuitionparadox}.  They extends straightforwardly to the case $\theta >1$, for which the unconditional mean waiting time $\langle t \rangle$ defined by (\ref{jhigkfmpwa}) in the limit $T \to +\infty$ is now finite and well-behaved.  This implies that $t_1 + t_2 +... + t_{K(t)}$ in (\ref{tgwtrgrwaaftr}) is well approximated by $K(t) \cdot \langle t \rangle$, which, being equal to $t$, yields $K(t) \sim t$, for all values of $\theta >1$. This extends the result (\ref{htyhtne})
previously derived only for $0 < \theta <1$.  Then, expression (\ref{h3iqpooadkv}) still holds and we obtain finally
that expression (\ref{hjuwosk}) holds with $K(t) \sim t$. For $n=1$, we recover that the leading term describing the time dependence of $\mathcal{A}(t)$ is a constant as described by (\ref{meanrasympmxleonepl}) for the case $\theta >1$. The simple scaling argument of subsection
\ref{intuitionparadox} shows that this result can be actually extended to all positive values of $\theta >1$.
For $n<1$ but $1-n$ small, $\mathcal{A}(t)\sim$ constant still holds for times $\ll 1/(1-n)$.

We can combine the results of the
intermediate asymptotics valid for $x \ll {1 \over (\theta-1)(1-n)}$ for $\theta\in(0,1)$ and for $\theta>1$ by
the following power law dependence of the mean activity
\begin{equation}\label{apowasallth}
\mathcal{A}(x,n) \sim x^{-\delta(\theta)}
\end{equation}
where the exponent $\delta(\theta)$ is given by
\begin{equation}\label{deltathdef}
\delta(\theta) = (1-\theta) \bs{1}(1-\theta) =
\begin{cases}
1-\theta~ , & 0<\theta<1~ , \\
0 ~ , & \theta>1~  .
\end{cases}
\end{equation}
Fig.~3 shows the exponent $\delta(\theta)$ given by \eqref{deltathdef} as a function of $\theta$.

\begin{quote}
\centerline{
\includegraphics[width=11cm]{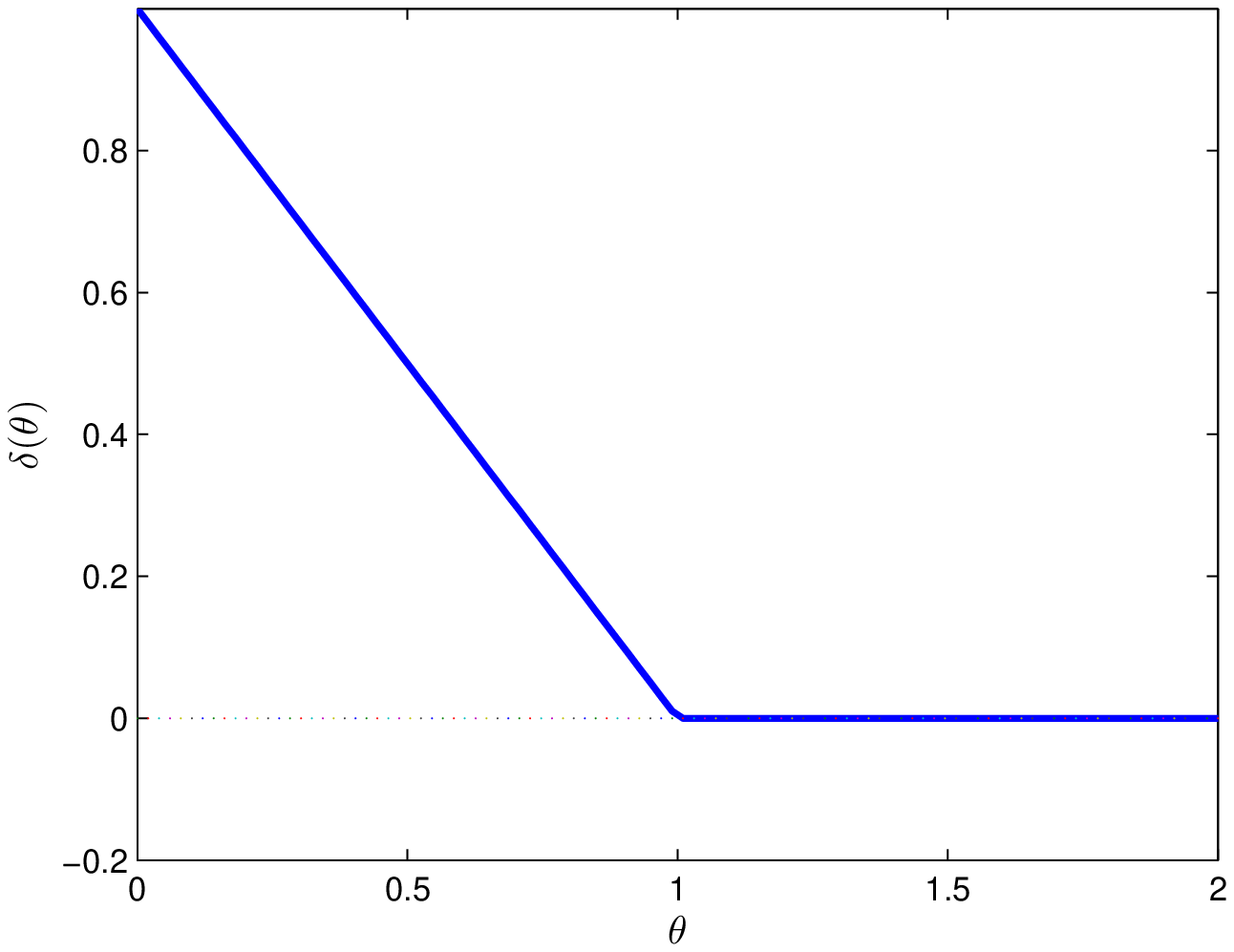}}
{\bf Fig.~3:} \small{Dependence of the exponent $\delta(\theta)$} given by (\ref{deltathdef}) of the mean activity $\mathcal{A}(x,n)$ as a function of $\theta$, in the intermediate power asymptotic defined  $1-n \ll 1$.
\end{quote}

\subsection{Long time asymptotic \label{tyhjrgbgrv}}

Let us now consider the asymptotic behavior of  integral \eqref{athmooneint} for $\varepsilon\gg 1$, where
$\varepsilon$ is defined in (\ref{epsdef}). In this case, the main contribution of the integral is the power law \eqref{powerlawstabdist} of the stable distribution, so that one may replace the function $\mathcal{A}_k(x;\theta)$ by
\begin{equation}\label{dtailasymp}
\mathcal{A}_k(x;\theta) \simeq {|\beta| \over |\Gamma(-\theta)|} ~ { k \over (x-|\alpha| k)^{\theta+1}} ~ .
\end{equation}
Accordingly, integral \eqref{athmooneint} takes the form
\begin{equation}\label{dinttailrestr}
\mathcal{A}_\text{int}(x,n) \simeq {|\beta| \over |\Gamma(-\theta)|}~ \int_1^{k_*-1} ~ { k~ e^{-\gamma k} ~ dk \over (x-|\alpha| k)^{\theta+1}} \qquad \varepsilon = {\gamma x \over |\alpha|} \gg 1 ~ .
\end{equation}
The upper limit of this integral removes the influence of the singularity at $k_*$ defined by (\ref{zeropointd})
which is irrelevant
for $\varepsilon\gg 1$. One may interpret the upper limit as resulting from  the short tail of the stable distribution.

Due to the fast decaying exponential $e^{-\gamma k}$, \eqref{dinttailrestr} can be approximated by
\begin{equation}\label{dinttailrest6j4ur}
\mathcal{A}_\text{int}(x,n) \simeq \left|{\beta \over \Gamma(-\theta)}\right|~ x^{-\theta-1} ~ \int_0^\infty ~  k~ e^{-\gamma k} ~ dk  \qquad \varepsilon = {\gamma x \over |\alpha|} \gg 1 ~ ,
\end{equation}
which finally leads to
\begin{equation}\label{dinttailrestyj3r}
\mathcal{A}_\text{int}(x,n) \simeq \left|{\beta \over \Gamma(-\theta)}\right|~ {1 \over \gamma^2} ~ x^{-\theta-1} \qquad \varepsilon  \gg 1 ~ ,
\end{equation}
If the kernel $\Phi(t)$ is of the form \eqref{powkern}, then $\beta= \Gamma(1-\theta)$ and one has
\begin{equation}\label{dinttailrestrqqq}
\mathcal{A}_\text{int}(x,n) \simeq {\theta \over \gamma^2} ~ x^{-\theta-1} \qquad \varepsilon  \gg 1 ~ .
\end{equation}

\subsection{Exact results for $\theta = 3/2$}

It is useful to check these results for $\theta=3/2$, for which we can make use of the explicit expression \eqref{stabthreetwo}. To be specific, we also assume that the kernel $\Phi(t)$ is given by formula \eqref{powkern}, so that the parameters $|\alpha|$ and $|\beta|$ are equal to
\begin{equation}
|\beta| = \left|\Gamma\left(-{1 \over 2}\right)\right| = 2 \sqrt{\pi}~ , \qquad |\alpha| = 2~ .
\end{equation}
Accordingly, the series in \eqref{atasympstabthmone} reads
\begin{equation}\label{atasympstabthnhrtwo}
\mathcal{A}(x,n) = \sum_{k=1}^\infty {n^k \over \sqrt[3]{4\pi}~ k^{2/3}} \psi\left( {x-2 k \over \sqrt[3]{4\pi}~ k^{2/3}};3/2 \right) ~ ,
\end{equation}
while the power asymptotics \eqref{meanrasympmxleonepl} and \eqref{dinttailrestr} take the form
\begin{equation}\label{bothasympththrtwo}
\mathcal{A}(x,n) \simeq
\begin{cases} \displaystyle
{1 \over 2} ~ , & \displaystyle {\gamma x \over 2} \ll 1~ ,
\\[4mm] \displaystyle
{3 \over 2 \gamma^2} ~ {1 \over x^{5/2} }~ , & \displaystyle {\gamma x \over 2} \gg 1~ .
\end{cases}
\end{equation}
Fig.~4 plots $\mathcal{A}(x,n)$ given by \eqref{atasympstabthnhrtwo} and its asymptotics given by\eqref{bothasympththrtwo}  as a function the reduced time $x$ for $n=0.99$. One can observe the predicted
constant plateau for times up to $\simeq 1/(1-n) \approx 10^2$, followed by the power law $\sim 1/t^{1+\theta}$
at larger times.

\begin{quote}
\centerline{
\includegraphics[width=11cm]{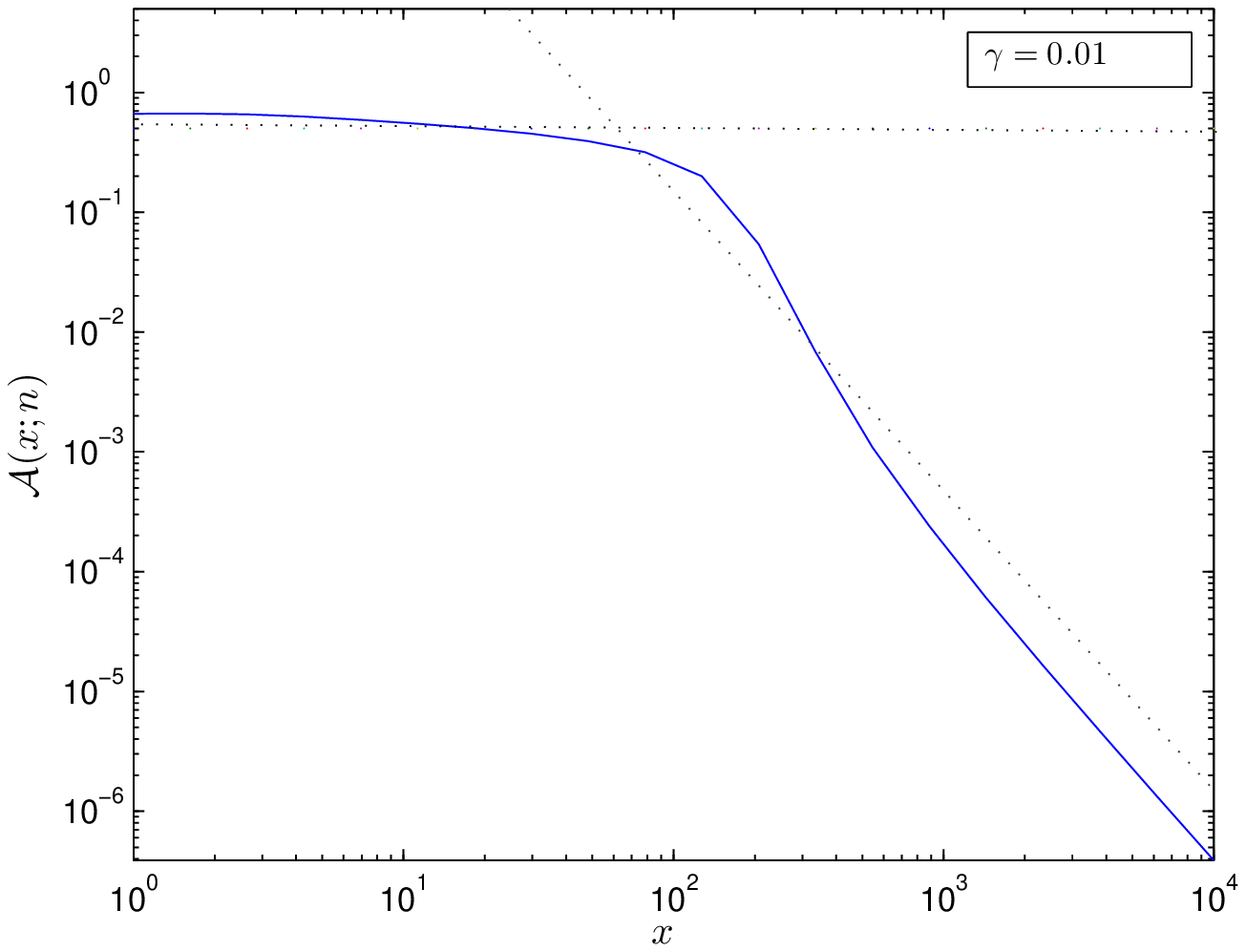}}
{\bf Fig.~4:} \small{Dependence of expression \eqref{atasympstabthnhrtwo}  for the mean activity (solid line) and its two asymptotic regimes  (\ref{bothasympththrtwo})
(dotted lines), for  $\theta=3/2$. The branching ratio is equal to $n=0.99$, i.e., $\gamma =0.01$.}
\end{quote}

\vskip 1cm
{\bf Acknowledgement}: We acknowledge financial support
from the ETH Competence Center "Coping with Crises in Complex
Socio-Economic Systems" (CCSS) through ETH Research Grant CH1-01-08-2.

\end{document}